\documentclass[twocolumn,preprintnumbers,amsmath,amssymb]{revtex4}
\usepackage{tabularx,graphicx}

\usepackage{color}
\usepackage{hyperref}
\hypersetup{
    colorlinks=true,
    linkcolor=blue,
    filecolor=blue,      
    urlcolor=blue,
}

\usepackage{color}

\usepackage{ulem}   

\begin{document}

\newcommand{\beq}{\begin{equation}}
\newcommand{\eeq}{\end{equation}}
\newcommand{\beqn}{\begin{eqnarray}}
\newcommand{\eeqn}{\end{eqnarray}}
\newcommand{\bmath}{\begin{subequations}}
\newcommand{\emath}{\end{subequations}}
\newcommand{\bra}[1]{\langle #1|}
\newcommand{\ket}[1]{|#1\rangle}

\title{Enormous variation in homogeneity and other anomalous features of room temperature superconductor samples: a Comment on Nature 615, 244 (2023)}

\author{J. E. Hirsch}
\address{Department of Physics, University of California, San Diego,
La Jolla, CA 92093-0319}

 \begin{abstract} 
  
The resistive transition width of a recently discovered room temperature near-ambient-pressure hydride superconductor \cite{roomtlh}
changes by more than three orders of magnitude between different samples, with the transition temperature nearly unchanged. For the narrowest transitions, the transition width relative to $T_c$ is only $0.014 \%$.
The voltage-current characteristics indicate vanishing critical current, and the normal state resistance is unusually small.
These anomalous behaviors and other issues indicate that this system is not a superconductor. Implications for other
hydrides are discussed.
\end{abstract}
\maketitle

 Room temperature at near ambient pressure has recently been claimed for a Lu-H-N compound \cite{roomtlh}.
 In this paper we question the validity ot that claim on the basis of analysis of some of the reported measurement results
 that show anomalous features, and discuss the implications of this analysis.
 
In Extended Data (hereafter ED) Fig. 15 of Ref. \cite{roomtlh}, the authors show resistance versus temperature in the absence and presence of a 
magnetic field. The relative width of the resistive transition $\Delta T/T_c$ shown in the inset of the figure is $0.13$ for zero field.
The authors explain the considerable width by stating {\it ``The large transition width at zero field indicates sample
inhomogeneities, which is typical for high-pressure experiments.''}, which is not implausible.

        \begin{figure} []
 \resizebox{8.5cm}{!}{\includegraphics[width=6cm]{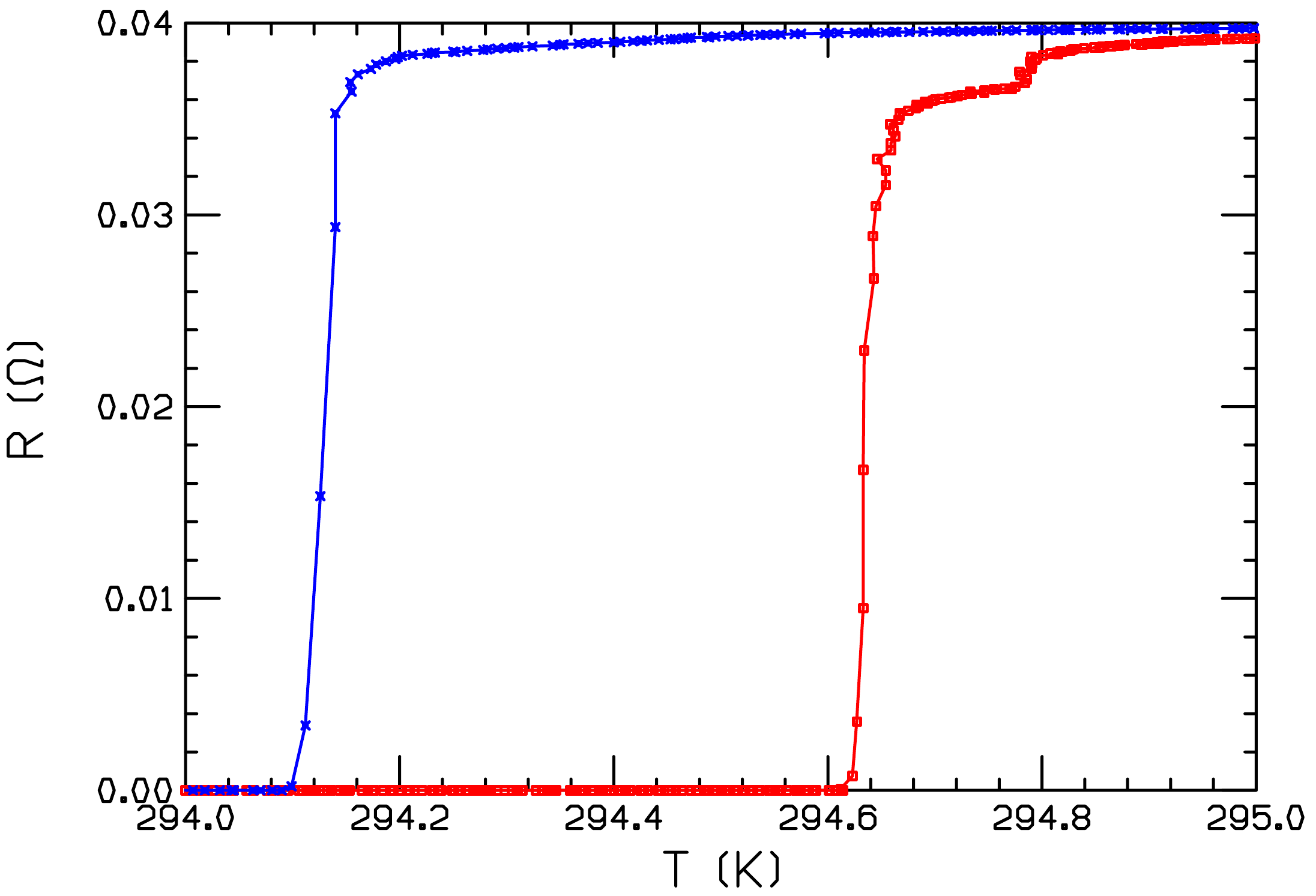}} 
 \caption {Resistance versus temperature  for N-doped  lutetium hydride at pressure $\sim 10 kbar$, as reported in 
 Ref. \cite{roomtlh} ED Fig. 13a and associated reported raw data. The blue and red curves were measured under cool down and warm up conditions
 respectively \cite{roomtlh}.}
 \label{figure1}
 \end{figure}

However, in the data for resistance versus temperature in ED Fig. 13a of Ref. \cite{roomtlh}, shown here in Fig. 1,  the width of the resistive transitions is only $0.04K$, so  the relative width is $\Delta T/T_c=0.00014$,
at comparable pressures (10 kbar vs 15 kbar respectively). For the resistance curves shown in Fig. 2
of Ref. \cite{roomtlh}, at pressures 10 kbar and 16 kbar, the relative width is
in-between those two extremes,  $\Delta T/T_c\sim 0.008$, 60 times larger
than for ED Fig. 13a, 16 times smaller than for ED Fig. 15.

Following the logic of the authors, the sample used for ED Fig. 13a is 1,000 times more homogeneous than the sample
used for ED Fig. 15, the samples used for their Fig. 2  are in-between. 
The protocol used in preparing these samples was presumably similar, as described in the ``Methods'' section of Ref. \cite{roomtlh}.
It is not understandable why superconducting samples prepared similarly would exhibit a degree of inhomogeneity that differs by
three orders of magnitude.  The fact that the transitions shown in Fig. 1 show hysteresis is also in conflict with what is expected for a superconductor
and is not explained in the paper.

        \begin{figure} []
 \resizebox{8.5cm}{!}{\includegraphics[width=6cm]{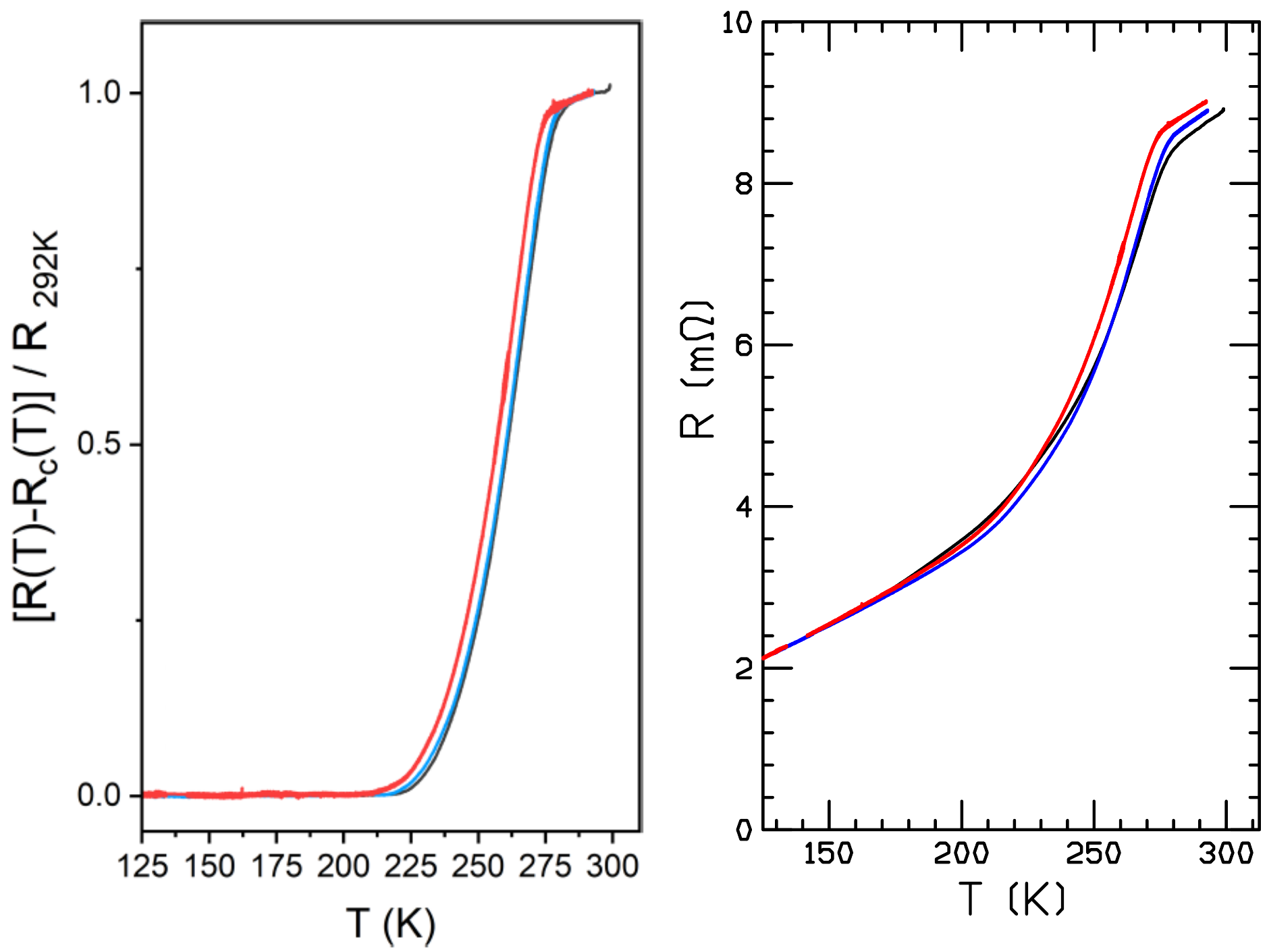}} 
 \caption {Resistance versus temperature  for N-doped  lutetium hydride at pressure $15 kbar$. Left panel:
 with background subtraction, as reported in  Ref. \cite{roomtlh} ED Fig. 15. Right panel:
 same without background subtraction, 
  obtained from the raw data given in 
 Ref. \cite{roomtlh}. The black, blue and red curves are for
 applied magnetic fields $0T$, $1T$, $3T$ respectively \cite{roomtlh}.}
 \label{figure1}
 \end{figure}

It should also be noted  that the authors say ``In some cases, small residual resistance
from the instrument offsets was subtracted from the measured voltage.'', but don't specify whether 
``some cases'' include the resistance data shown in their Fig. 2 and ED Fig.13a. For ED Fig. 15 it is explicitly stated that
a background resistance is subtracted out \cite{roomtlh}. When plotting the raw data without background
subtraction, the curves shown in Fig. 2 right panel result \cite{bar}. There is no hint of a superconducting transition in Fig. 2
right panel.

  \begin{figure} [t]
 \resizebox{8.5cm}{!}{\includegraphics[width=6cm]{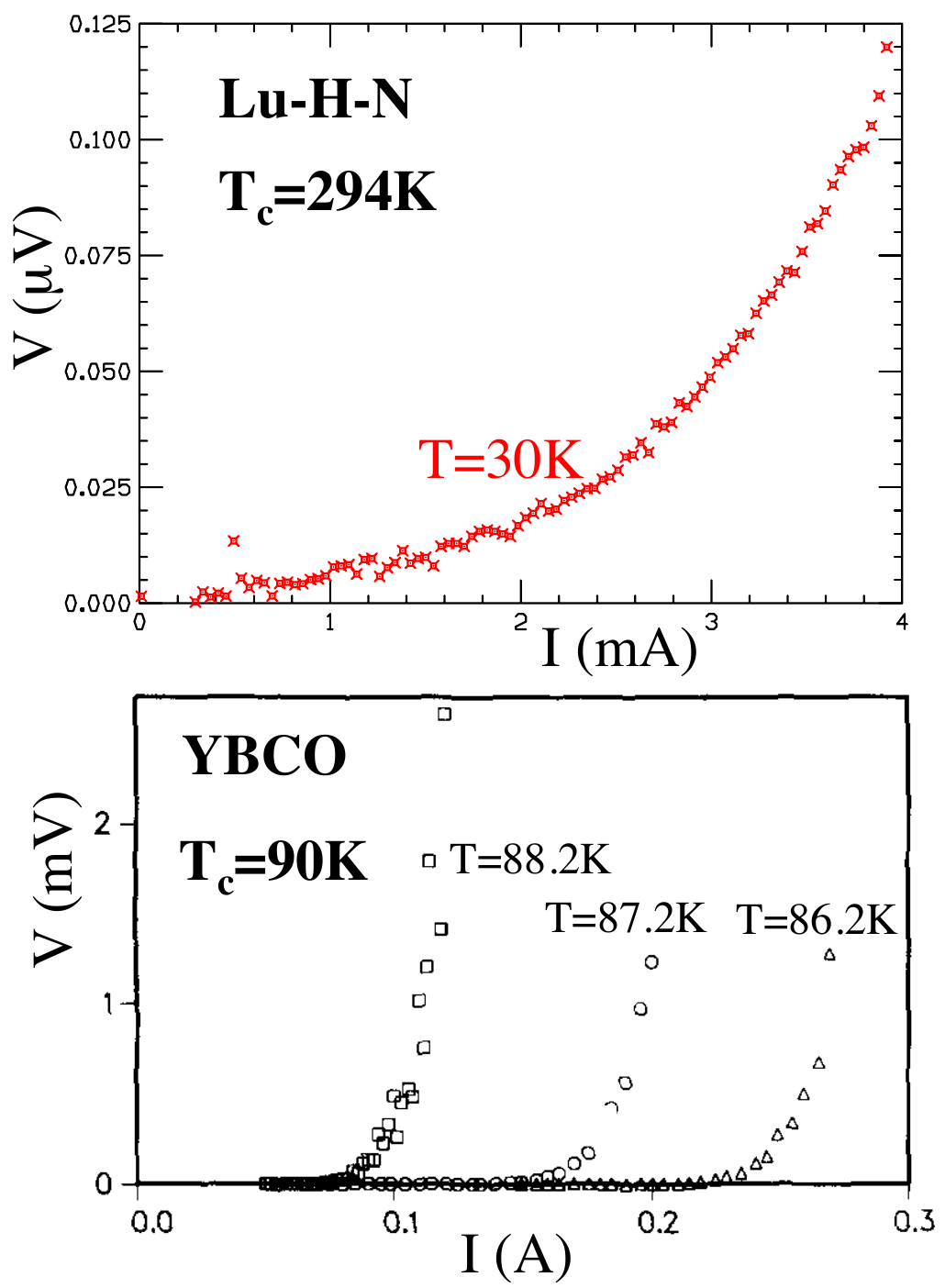}} 
 \caption {Top panel: voltage versus current from source data for Fig. 2b of Ref. \cite{roomtlh}.
 Bottom panel: voltage versus current data for a known superconductor, from Ref. \cite{visuper}.}
 \label{figure1}
 \end{figure}

In addition, the width of the transitions shown in Fig. 1 is unreasonably small. No other known superconductor exhibits such
sharp transitions even at ambient pressure, and under pressure additional broadening of the transition  results from
pressure gradients. In ref. \cite{hmnature}, we pointed out that the narrow width of the transitions reported in Ref. \cite{roomt} for another
room temperature superconductor under pressure,  CSH, was unreasonably small, $\Delta T/T_c=0.005$, and that was 35 times $larger$ than the width seen
in Fig. 1. Other anomalies in the resistance curves of Ref. \cite{roomt} were  noted in Ref. \cite{hamlin}. Ref. \cite{roomt}, which has
six coauthors in common with Ref. \cite{roomtlh}, was recently retracted \cite{roomtr}.


Furthermore, the voltage-current characteristic shown in Fig. 2b of Ref. \cite{roomtlh}, reproduced here in Fig. 3 top panel, is not consistent with what is expected for a superconductor at temperature well below its critical temperature. There is no evidence of any
region of zero resistance in Fig. 3  top panel, in contrast with the typical behavior shown in Fig. 3 bottom panel for a known superconductor \cite{visuper}, showing
zero voltage up to a critical current that increases as the temperature decreases.
Other such measurements for known superconductors showing how this behavior varies with temperature and magnetic field
are shown   in Refs. \cite{visuper2,visuper3,visuper4,visuper5}.
If one nevertheless insisted to infer a non-zero critical current from Fig. 3 top panel, it would be certainly smaller than $0.5mA$,  at temperature $T/T_c\sim0.1$. According to the authors, ``On average, sample sizes are on the
order of 70-100 $\mu m$ in diameter and 10-20 $\mu m$ thick''. Assuming the smallest cross-sectional area  in that range, $A\sim 10\mu m \times 70\mu m$,
yields a critical current density $J_c<72A/cm^2$ (or smaller if larger dimensions are assumed). That would be five orders of magnitude smaller than critical current densities reported
for $LaH_{10}$ and $H_3S$ at temperature $T/T_c\sim 0.5$  \cite{e2021p}, and inconsistent with the magnetization measurements shown in
Fig. 3 of Ref. \cite{roomtlh}.

  \begin{figure} [t]
 \resizebox{8.5cm}{!}{\includegraphics[width=6cm]{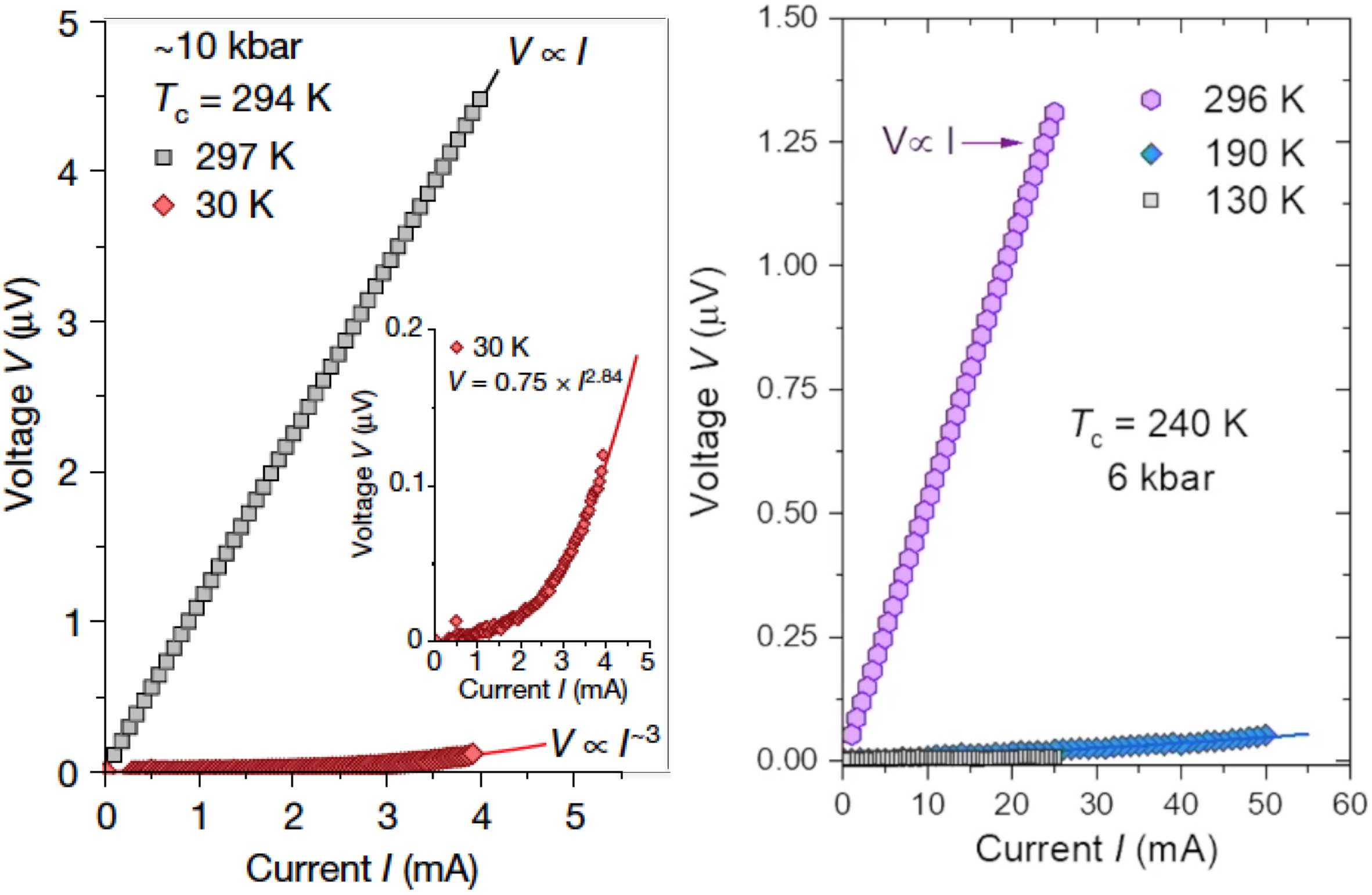}} 
 \caption {Voltage versus current for Lu-H-N. Left panel:  Fig. 2b of Ref. \cite{roomtlh}. Right panel:
 another example reported by R. P. Dias, Ref. \cite{apstalk}.}
 \label{figure1}
 \end{figure}

Furthermore,  the values of the normal state resistance at room temperature inferred from Fig. 2b of Ref. \cite{roomtlh},
and for another example of voltage current-characteristics presented by the lead author of Ref. \cite{roomtlh} in Ref. \cite{apstalk}, shown on the left and
right panels of Fig. 4, are anomalously small: $R\sim 1.12 m\Omega$ for the left panel, $R\sim 0.05 m\Omega$ for the right panel. From the van der Pauw formula $\rho\sim \pi d R /ln(2)$  (d=thickness, R=resistance), assuming the largest $d=20\mu m$   yields $\rho \sim 10 \mu \Omega -cm$
for the left panel of Fig. 4 and  $\rho \sim 0.4 \mu \Omega -cm$
for the right panel. These values are six times and 130 times smaller that the room temperature resistivity of 
lutetium metal. These discrepancies point to the behavior reported \cite{roomtlh,apstalk} as voltage versus current for Lu-H-N both at room temperature
and at low temperatures   being due to experimental artifacts rather than superconductivity.
It is notable that for other samples that showed a resistance drop from much larger room temperature resistance values
(Fig. 2a and ED Fig. 13 a of Ref. \cite{roomtlh}), consistent with what could be 
expected, no voltage-current characteristics were shown in Refs. \cite{roomtlh,apstalk}.

   \begin{figure*} [t]
 \resizebox{16.5cm}{!}{\includegraphics[width=6cm]{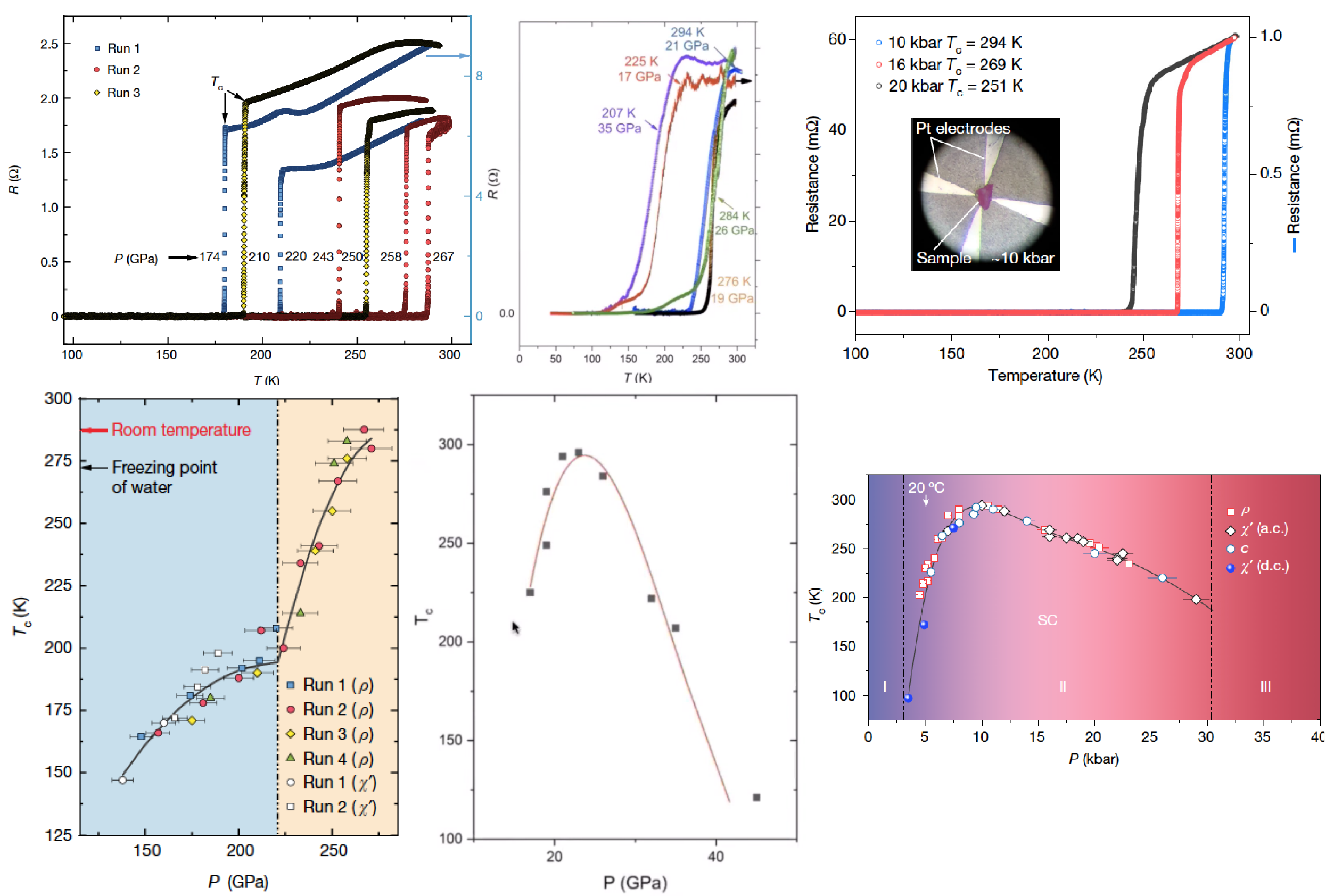}} 
 \caption {Three holy grails. The top three panels show resistance versus temperature for three different compounds at pressure ranges
 $\sim 200 GPa$ (Refs. \cite{roomt,roomt2,roomt3}), $\sim 20 GPa$ (Ref. \cite{one}) and $\sim 2GPa$ (Ref. \cite{roomtlh}) respectively, all showing room temperature superconductivity.
 The bottom three panels show $T_c$ versus pressure for the three different compounds as reported in Refs.  \cite{roomt}, \cite{one}, \cite{roomtlh}.
 Note that even though Ref. \cite{roomt} was retracted \cite{roomtr}, all the authors disagreed with the retraction.}
 \label{figure1}
 \end{figure*}

 We also point out that the ac susceptibility data shown in ED Fig. 5 of Ref. \cite{roomtlh} before background
 subtraction  
 show a background dependence on temperature that has positive slope, negative slope, and zero
 slope, for the same or comparable pressures. The background ac susceptibility  is expected to reflect
 the physical properties of the environment of the sample, which should not drastically change 
 for different measurements. 
 
 
 
   \begin{figure*} [t]
 \resizebox{16.5cm}{!}{\includegraphics[width=6cm]{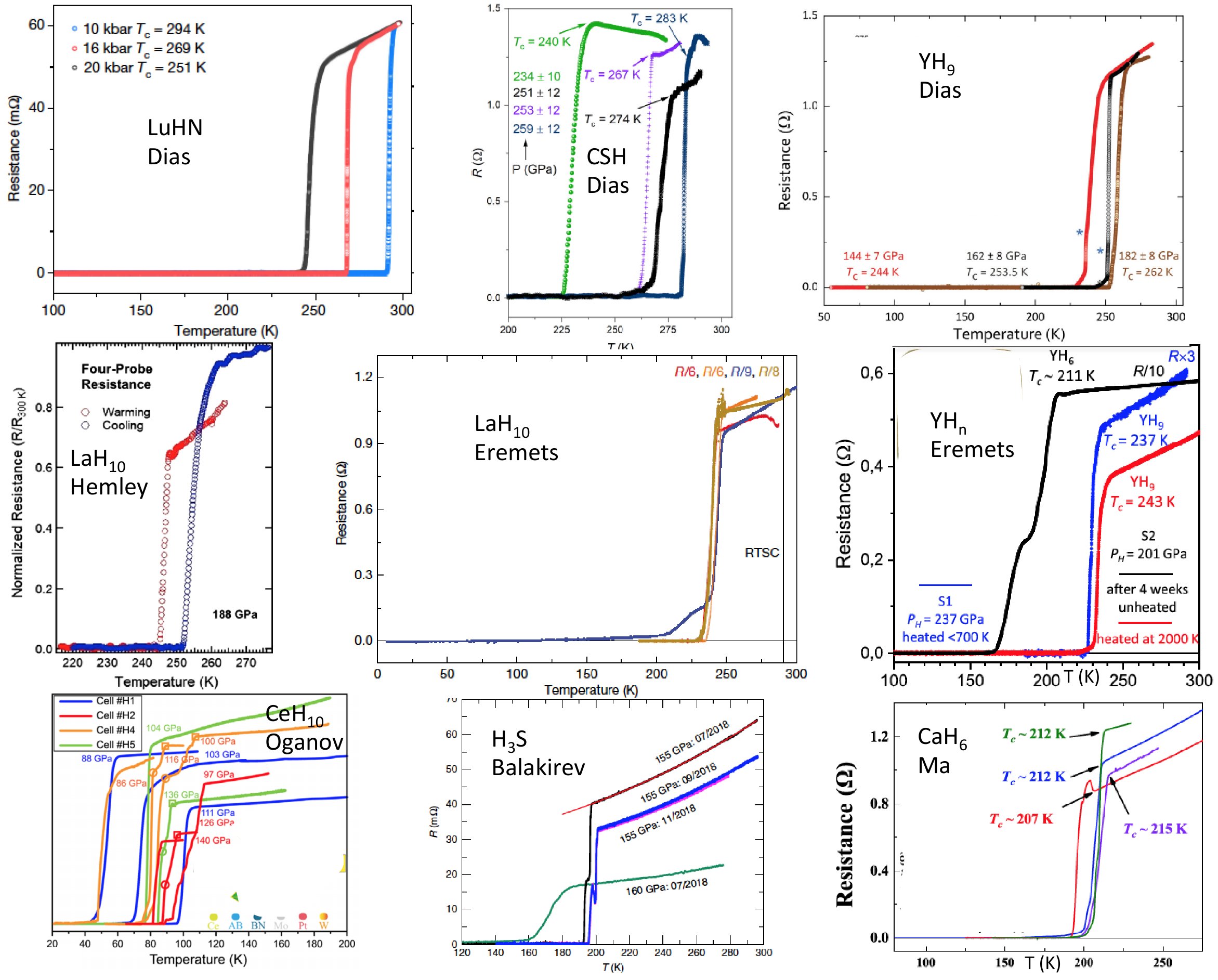}} 
 \caption {Resistance versus temperature for various hydrides under pressure claimed to be high temperature superconductors. 
The name of the last author in the reference is shown in each panel. From upper left to
 lower right: $LuHN$ \cite{roomtlh},$CSH$ \cite{roomt}, $YH_9$ \cite{ydias}, $LaH_{10}$ \cite{hemley}, $LaH_{10}$ \cite{laeremets},
 $YH_6$ \cite{yeremets},
 $CeH$ \cite{ceh}, $H_3S$ \cite{e2019}, $CaH_6$ \cite{cah6}.}
 \label{figure1}
 \end{figure*}

We also  point out that the lead author of Ref. \cite{roomtlh}  R. P. Dias and some of its coauthors have 
previously reported room temperature superconductivity in other compounds,
at pressures one \cite{one} and two \cite{roomt,roomt2,roomt3} orders of magnitude larger than reported in Ref. \cite{roomtlh},
as shown in Fig. 5.
In the 112 years since superconductivity was discovered,
no room temperature superconductivity has been conclusively established by other researchers
in any compound at any pressure despite intensive  searches.
The probability that the same research group would hit this holy grail three separate  times is insignificant.

Finally, we point out that experimental attempts to reproduce the results reported in Ref. \cite{roomtlh} 
 have shown no indication of superconductivity in samples prepared  by following the sample preparation method described in Ref. \cite{roomtlh} \cite{rep1,rep2,rep3,rep4,rep5,rep6}, and theoretical attempts to calculate $T_c$ in this system
 within the conventional theory of superconductivity have found values of $T_c$ two orders of magnitude
 smaller \cite{repth,repth2,repth3,repth4} than reported in Ref. \cite{roomtlh}.

In conclusion, the extreme sharpness of the resistive transition curves shown in Fig. 1,  the fact that the width of the 
resistive transition changes
by three orders of magnitude between different samples, the fact  that resistance data versus temperature plotted
without background subtraction show no hint of superconductivity,  the fact that  voltage-current characteristics  do not  show evidence for a finite critical current, the fact that normal state resistances measured are anomalously small, the fact that the background ac
susceptibility changes drastically in different measurements, 
and  the fact  that several experimental and theoretical studies have not been able to reproduce the
results reported in Ref. \cite{roomtlh}, indicate that the behavior observed reported in Ref. \cite{roomtlh} is not due to superconductivity.

If the resistance drops seen in Fig. 5 and  other figs.
of Refs. \cite{roomtlh,roomt,roomt2,roomt3} are not due to superconductivity they must be due to other reasons 
unrelated to superconductivity, either physical phenomena or/and experimental artifacts associated with performing resistance measurements on  very small hydrogen-rich samples under high pressure in diamond-anvil cells, with the expectation grounded in the conventional
theory of superconductivity that superconductivity will be found
\cite{pickett}.
The resistance versus temperature curves shown in Fig. 5 above and in other figs.
of Refs. \cite{roomtlh,roomt,roomt2,roomt3} look similar to resistance versus  temperature  curves for other hydrides under high pressure
that have been claimed to be high temperature superconductors in recent years \cite{troyan22}, as shown in Fig. 6. This raises the possibility that
those same other reasons unrelated to superconductivity that may account for the resistance drops in  Refs. \cite{roomtlh,roomt,roomt2,roomt3} 
reported by Dias and
coworkers could account for the resistance drops seen in
all hydrides under high pressure claimed to be high temperature
superconductors \cite{troyan22}, in contradiction with the predictions of Ref. \cite{pickett} and in agreement with other theoretical predictions \cite{xor}.
Magnetic evidence claimed to support high temperature superconductivity in sulfur and lanthanum hydrides under pressure has been called into
question elsewhere \cite{mag1,hmmre,huangmine}.
\newline
 \begin{acknowledgments}
The author is grateful  to multiple colleagues for discussions on these issues, and in particular to Erik van Heumen for stimulating 
discussions on   transport measurements. Some of the anomalies reported here were also noted in comments to Ref. \cite{roomtlh} submitted by readers that are posted at the website of Ref. \cite{roomtlh}, at pubpeer.com \cite{pubpeer}, in the blog post ``nanoscale views'' \cite{nano},  and in various postings in reddit.com \cite{reddit}.
\end{acknowledgments}
\hfill \break
\noindent {\bf Competing interests:} the author declares no competing interests.
\hfill \break
\noindent {\bf Data availability statement:} The data that support the findings of this study are available from the author upon  reasonable request.

 \end{document}